\documentclass[12pt]{article}
\usepackage{graphicx}
\usepackage{amsmath}
\usepackage{xspace}
\usepackage{overpic}

\newcommand\pubnumber{WSU--HEP--XXYY}
\newcommand\pubdate{\today}

\def\wayne{School of Physics and Astronomy\\
The University of Manchester, M13 9PL, UK}

\def\Title#1{\begin{center} {\Large #1 } \end{center}}
\def\Author#1{\begin{center}{ \sc #1} \end{center}}
\def\Address#1{\begin{center}{ \it #1} \end{center}}

\newcommand\pubblock{\rightline{\begin{tabular}{l} \pubnumber\\
         \pubdate  \end{tabular}}}
\newenvironment{Abstract}{\begin{quotation}  }{\end{quotation}}
\newenvironment{Presented}{\begin{quotation} \begin{center} 
             PRESENTED AT\end{center}\bigskip 
      \begin{center}\begin{large}}{\end{large}\end{center} \end{quotation}}

\def\beq{\begin{equation}}
\def\eeq#1{\label{#1}\end{equation}}
\def\eeqn{\end{equation}}

\def\beqa{\begin{eqnarray}}
\def\eeqa#1{\label{#1}\end{eqnarray}}
\def\eeqan{\end{eqnarray}}

\let\bar=\overbar

\def\Dslash{\not{\hbox{\kern-4pt $D$}}}
\def\dslash{\not{\hbox{\kern-2pt $\del$}}}

\def\msb{{\bar{\ssstyle M \kern -1pt S}}}

\newcommand{\tmmm}{\ensuremath{\tau^-\to \mu^+\mu^-\mu^-}\xspace}
\newcommand{\tpmm}{\ensuremath{\tau\to p\mu\mu}\xspace}
\newcommand{\tpmmOS}{\ensuremath{\tau^-\to \bar{p}\mu^+\mu^-}\xspace}
\newcommand{\tpmmSS}{\ensuremath{\tau^-\to p\mu^-\mu^-}\xspace}
\newcommand{\DsPhiPi}{\ensuremath{D_s^-\to\phi (\mu^+\mu^-) \pi^-}\xspace}

\newcommand{\tev}{\ensuremath{\mathrm{\,Te\kern -0.1em V}}\xspace}
\newcommand{\DsEtaMuNu}{\ensuremath{D_s^- \to \eta (\mu^+\mu^-\gamma) \mu^- \nu_{\bar{\mu}}}\xspace}
\newcommand{\DsTauNu}{\ensuremath{D_s^-\to \tau^- \nu_{\bar{\tau}}}\xspace}
\newcommand{\gl}{\ensuremath{{\rm \mathcal{M}_{3body}}}\xspace}
\newcommand{\pid}{\ensuremath{{\rm \mathcal{M}_{PID}}}\xspace} 
\newcommand{\Jpsimumu}{\ensuremath{J/\psi\to \mu^+\mu^-}\xspace}

\def\BF         {{\ensuremath{\cal B}}}

\begin{document}
\begin{titlepage}
\pubblock

\vfill
\Title{$\tau$ physics at LHCb}
\vfill
\Author{Jon Harrison}
\Address{\wayne}
\vfill
\begin{Abstract}
We report on the first searches for lepton flavour violating $\tau^-$ decays at a hadron collider. These include 
searches for the lepton flavour violating decay \tmmm and the lepton flavour
and baryon number violating decays \tpmmOS and \tpmmSS. Upper limits of
$\BF(\tmmm) < 4.6 \times 10^{-8}$, $\BF(\tpmmOS) < 3.4 \times 10^{-7}$ and
$\BF(\tpmmSS) < 4.6 \times 10^{-7}$ are set at 90\% confidence level. 
A measurement of the inclusive $Z\to\tau^+\tau^-$ cross-section at 7\,TeV is also reported and is found 
to be consistent with the Standard Model. The ratio of the $Z\to\tau^+\tau^-$ cross-section to the $Z\to\mu^+\mu^-$ cross-section is found to be consistent with lepton universality.

\end{Abstract}
\vfill
\begin{Presented}
The 7th International Workshop on Charm Physics (CHARM 2015)\\
Detroit, MI, 18-22 May, 2015
\end{Presented}
\vfill
\end{titlepage}
\def\thefootnote{\fnsymbol{footnote}}
\setcounter{footnote}{0}
%

\section{$\tau$ lepton flavour and baryon number violation}
\subsection{Introduction}
\label{ssec:intro}

Lepton flavour violating (LFV) processes are allowed within the context of the Standard Model (SM) with massive neutrinos, 
albeit with vanishingly small branching fractions. Many New Physics (NP) models allow for enhanced rates which approach 
current experimental sensitivities in certain regions of parameter space \cite{lfvreview}. 
Setting limits on the branching fractions of such decays helps to constrain these models, whilst a direct observation 
of charged LFV would be a clear sign of NP.

At the LHCb experiment~\cite{lhcb}, the inclusive $\tau^-$ production cross-section\footnote{The inclusion of charge conjugate processes is implied throughout.} is large, such that in one nominal year of data taking 
a factor of $\sim100$ more $\tau^-$ leptons are produced than in the total samples collected over the lifetimes of the BaBar and Belle experiments.   
This, combined with the clean detector signatures provided by final-state muons, makes searches for decays such as \tmmm particularly 
promising. The current world's best limit on the branching fraction for \tmmm is $2.1\times10^{-8}$ at 90\% confidence level (CL) from Belle~\cite{belle}.

The physics reach of these searches can be further extended by considering decays such as \tpmmOS and \tpmmSS, 
which are also baryon number violating and lepton number violating, 
whilst still containing multiple final-state muons. These decays both have $\Delta( B - L ) = 0$ ($B$ and $L$ are the net baryon number and lepton number respectively), 
as required by most extensions 
of the SM, but could have rather different NP interpretations~\cite{Raidal}. No measurements for these decays currently exist, but complementary 
searches such as $\tau^- \rightarrow \Lambda h^-, \bar{\Lambda} h^-$ (with $h = \pi,K$) have been performed by 
BaBar and Belle, with limits of order $10^{-7}$ obtained~\cite{hfag}. 

In the following we describe the searches for \tmmm, \tpmmOS and \tpmmSS decays~\cite{tmmmOld, tmmm} using samples of proton-proton collisions collected at $\sqrt{s} = 7\tev$ in 2011 
and $\sqrt{s} = 8\tev$ in 2012. For the \tpmmOS and \tpmmSS analyses only the 7\tev dataset is used, which corresponds to an integrated luminosity of 1.0\,fb$^{-1}$. 
For the \tmmm analysis the addition of the 8\tev sample gives a combined integrated luminosity of 3.0\,fb$^{-1}$ and supercedes the results from Ref.~\cite{tmmmOld}. 

\subsection{Event selection}
\label{ssec:sel}

After passing the LHCb trigger~\cite{trigger}, candidates are selected with loose cuts 
based on the kinematics of the reconstructed particles. For the \tpmmOS and \tpmmSS channels 
loose particle identification (PID) requirements are also applied. As the analysis is 
performed blind, initially candidates within $\pm$30 MeV/$c^2$ of $m_{\tau}$ are excluded.

\begin{figure}[!t]
\centering
\includegraphics[width=0.45\textwidth]{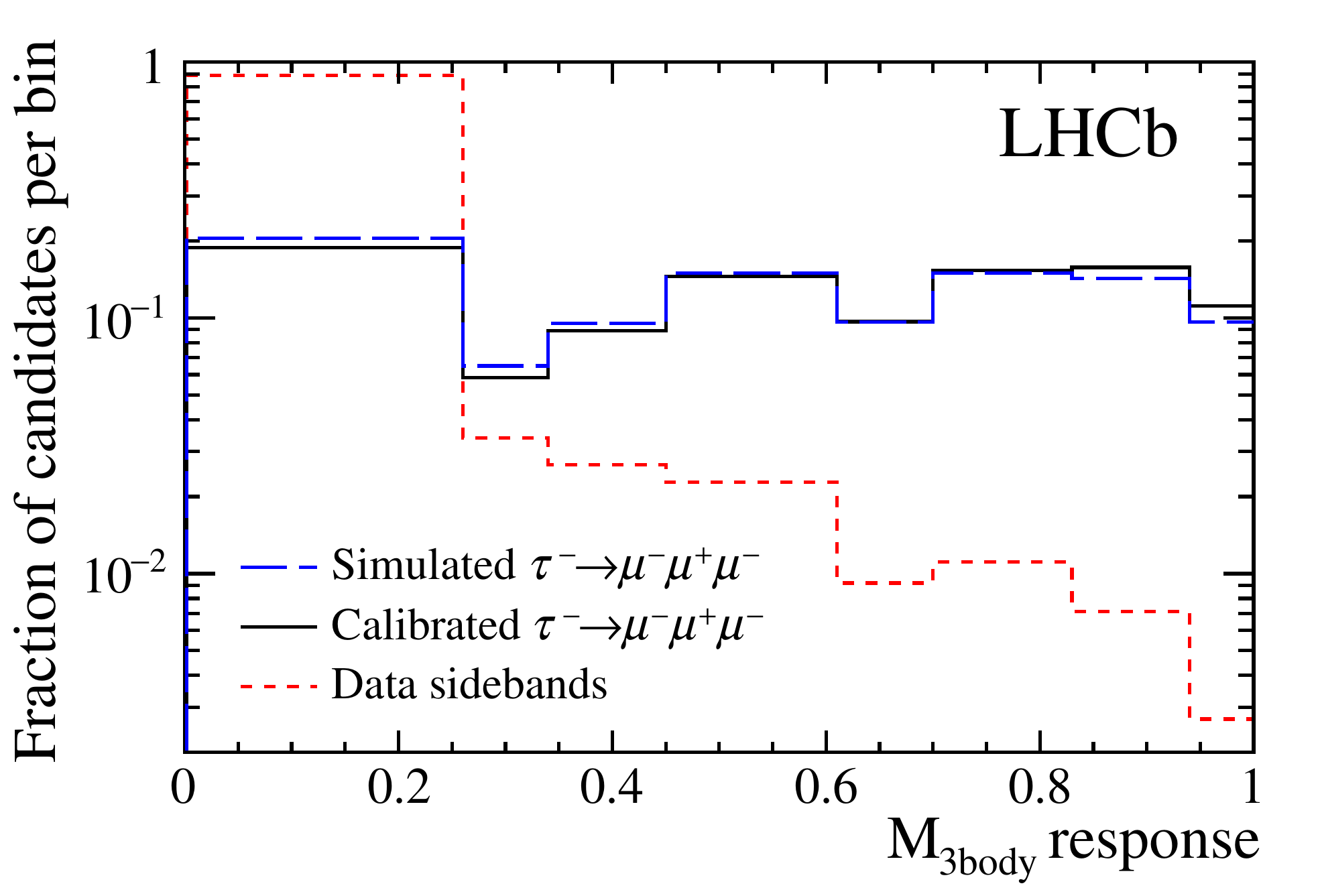}
\includegraphics[width=0.45\textwidth]{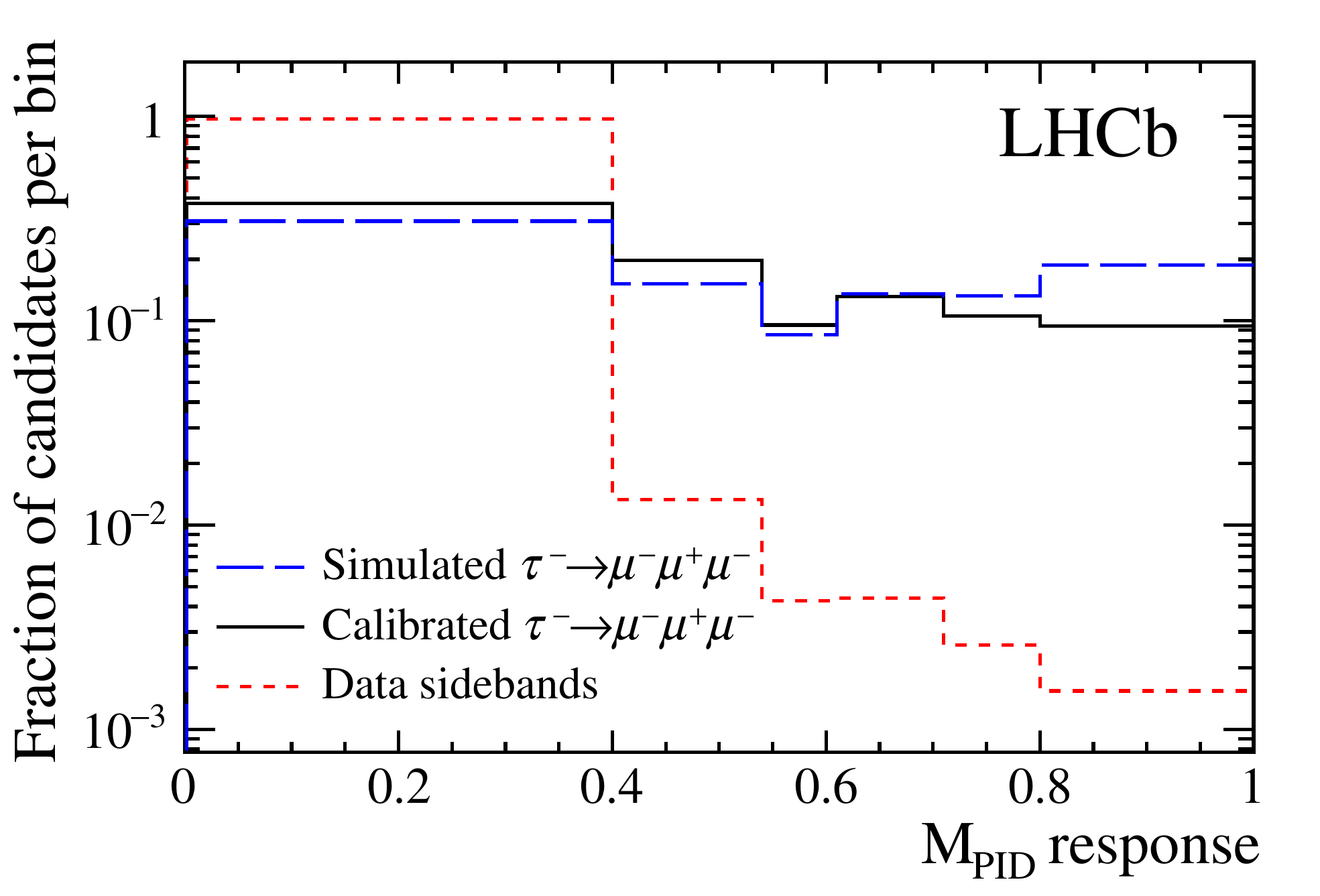}
\caption{Response of the \gl (left) and \pid (right) likelihoods for \tmmm signal MC (solid and long-dashed lines) and data sideband (short-dashed lines) candidates.}
\label{fig:gl}
\end{figure}

\begin{figure}[!t]
\centering
\includegraphics[width=0.5\textwidth]{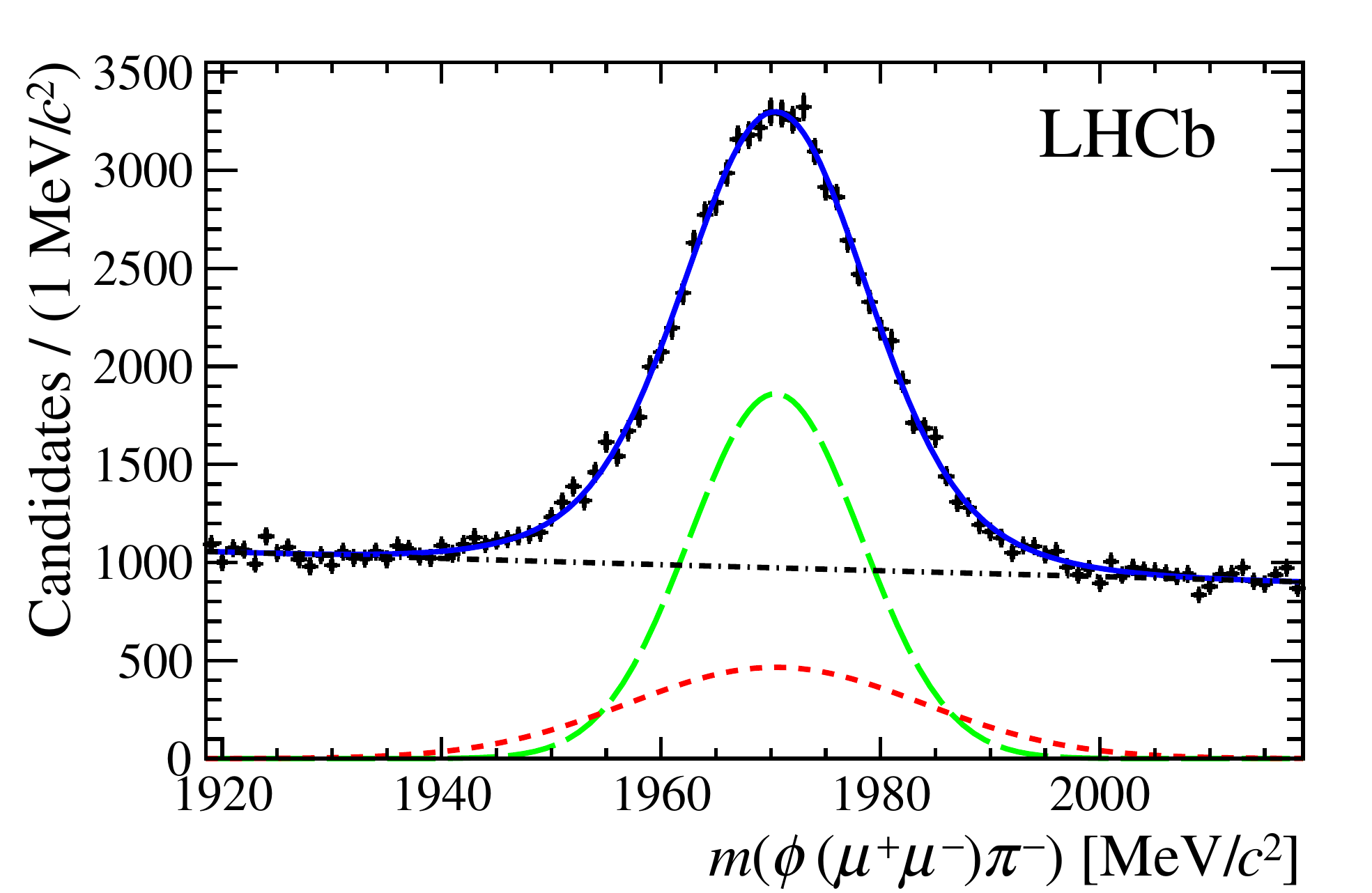}
\caption{Invariant mass distribution of \DsPhiPi candidates in 8 TeV data. The solid line shows the overall fit,
the long-dashed and short-dashed lines show the two Gaussian components of the $D^-_s$ signal and the dot-dashed line shows the combinatorial
background contribution.}
\label{fig:ds}
\end{figure}

In the case of \tmmm, the candidates are then classified according to three likelihoods, \gl, \pid 
and the reconstructed invariant mass of the $\tau^-$ candidate. The multivariate classifier \gl uses the 
kinematic and geometrical properties of the $\tau^-$ candidate to distinguish displaced 3-body 
decays from $N$-body ($N > 3$) and combinations of tracks from different vertices. 
The multivariate classifier \pid quantifies the compatibility of each of the three decay products 
with the muon hypothesis, using information from the Ring Imaging Cherenkov detectors, calorimeters and muon stations.
Both classifiers are trained on signal and background (inclusive $b\bar{b}$ and inclusive $c\bar{c}$) 
Monte Carlo (MC) and are calibrated on \DsPhiPi and \Jpsimumu data respectively. The data is binned in bins of \gl and \pid, and the number of bins 
and the position of the bin boundaries in each classifier are optimised, depending on the classifier and the data taking year. 
Figure~\ref{fig:gl} shows the response of \gl and \pid for \tmmm signal MC and data outside of the signal region, which is referred to in the following as the data sidebands. 
For invariant mass classification the signal shape in the mass window is taken from a fit 
to \DsPhiPi events in data, shown in Figure~\ref{fig:ds} for 8\tev data with the \tmmm selection. 
Both the central value of the mass window and the mass resolution are corrected 
according to the measured scaling and resolution in data at LHCb.


For \tpmmOS and \tpmmSS the candidates are classified according to \gl and the invariant mass 
of the $\tau^-$ candidate, in an identical manner to \tmmm. Particle identification requirements are then imposed through the application of hard cuts 
on the inputs to \pid to separate protons from pions and muons from charged hadrons.

\subsection{Normalisation}
\label{ssec:norm}

The observed number of \tmmm, \tpmmOS or \tpmmSS candidates is converted into a branching fraction by 
normalising to the \DsPhiPi calibration channel according to
\begin{align}
{\BF(\tau\to X\mu\mu)} = 
\quad {\BF(D_{s}\rightarrow\phi(\mu\mu)\pi)}
\times
\frac{f_{\tau}^{D_{s}}}{\BF(D_{s}\rightarrow\tau\nu_{\tau})}
\times
\frac{\rm
{\epsilon\mathstrut_{D_{s}\rightarrow\phi\pi}}
}{\rm
{\epsilon\mathstrut_{\tau\rightarrow X\mu\mu}}
}
\times\frac{N_{\tau\rightarrow X\mu\mu}}{N_{D_{s}\rightarrow\phi\pi}}.\nonumber
\label{eq:normalization}
\end{align}
The branching fraction of \DsPhiPi is determined from known branching fractions taken from Ref.~\cite{pdg}. 
The branching fraction of \DsTauNu is also taken from Ref.~\cite{pdg}.
The quantity $f_{\tau}^{D_{s}}$ is the fraction of $\tau^-$ that come from $D_s^-$ decays, calculated using the $b\bar{b}$ and $c\bar{c}$ 
cross-sections as measured by LHCb~\cite{bCrossSec8TeV, sigmaccLHCb}, the inclusive $b\to\tau^-$ and $c\to\tau^-$ 
rates as measured by the LEP experiments and various branching fractions from Ref.~\cite{pdg}. 
This term is required as \DsTauNu only accounts for $\sim80\%$ of the production of $\tau^-$ leptons. 
The total efficiencies, $\epsilon_{\tau\rightarrow X\mu\mu}$ and $\epsilon_{D_{s}\rightarrow\phi\pi}$, take 
into account the acceptance, selection and trigger efficiencies for the signal and normalisation channels, 
whilst $N_{D_{s}\rightarrow\phi\pi}$ is the number of reconstructed \DsPhiPi events in data, and $N_{\tau\rightarrow X\mu\mu}$ is the observed number of signal events. 

For \tpmmOS and \tpmmSS the normalisation is identical, aside from the inclusion of the PID cut 
efficiencies in $\epsilon_{\tpmm}$ and $\epsilon_{D_{s}\rightarrow\phi\pi}$. The tight PID requirements result in a larger normalisation factor, but a reduced 
background level, such that the overall effect is similar to the use of the multivariate classifier.

The advantage of the relative normalisation is that many of the systematic errors that are common to both the signal and normalisation channels 
cancel and an explicit knowledge of the luminosity and the inclusive $\tau^-$ cross-section are not required.

\subsection{Results}
\label{ssec:results}

\begin{figure}[!t]
\centering
\includegraphics[width=0.41\textwidth]{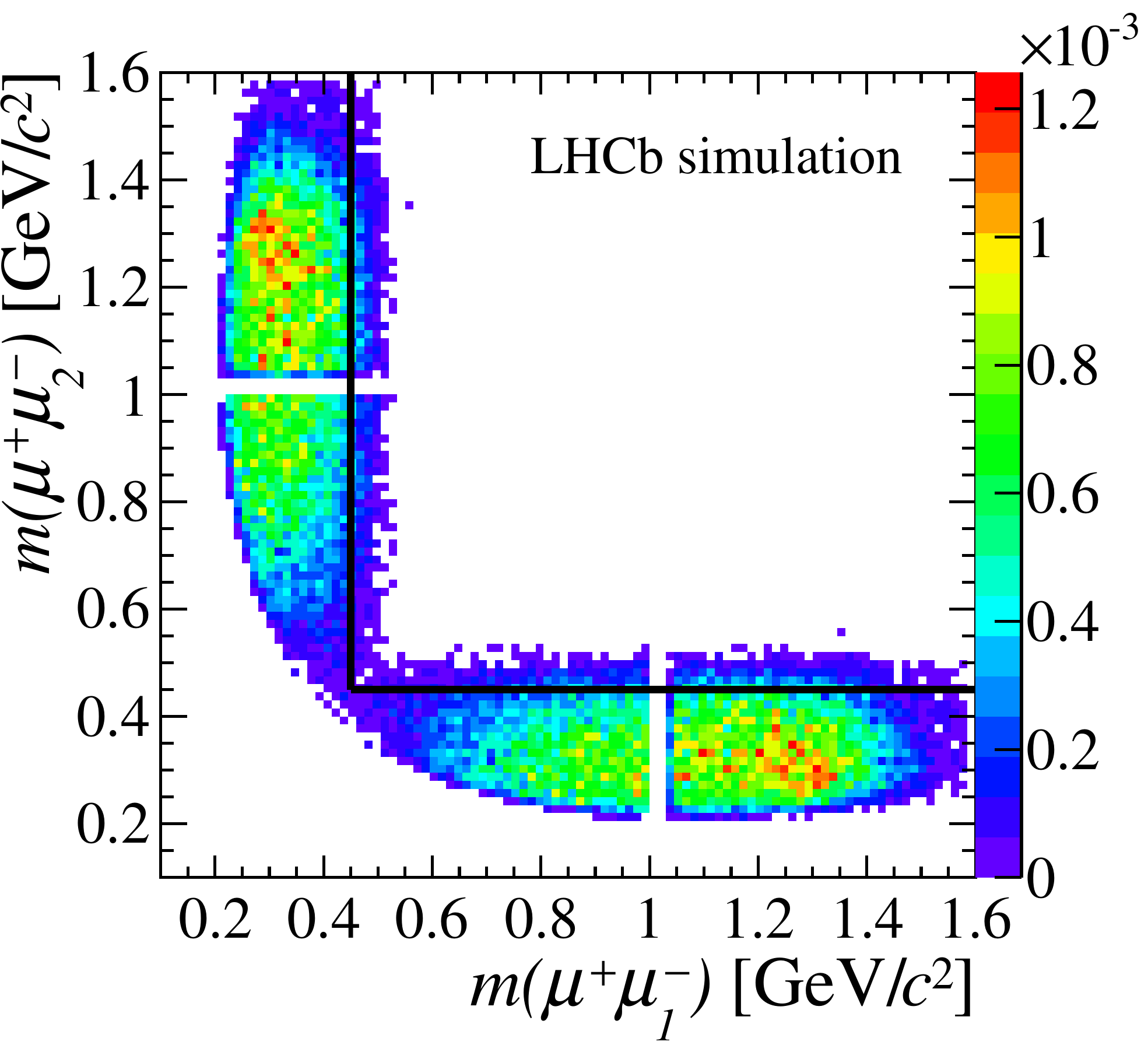}
\includegraphics[width=0.53\textwidth]{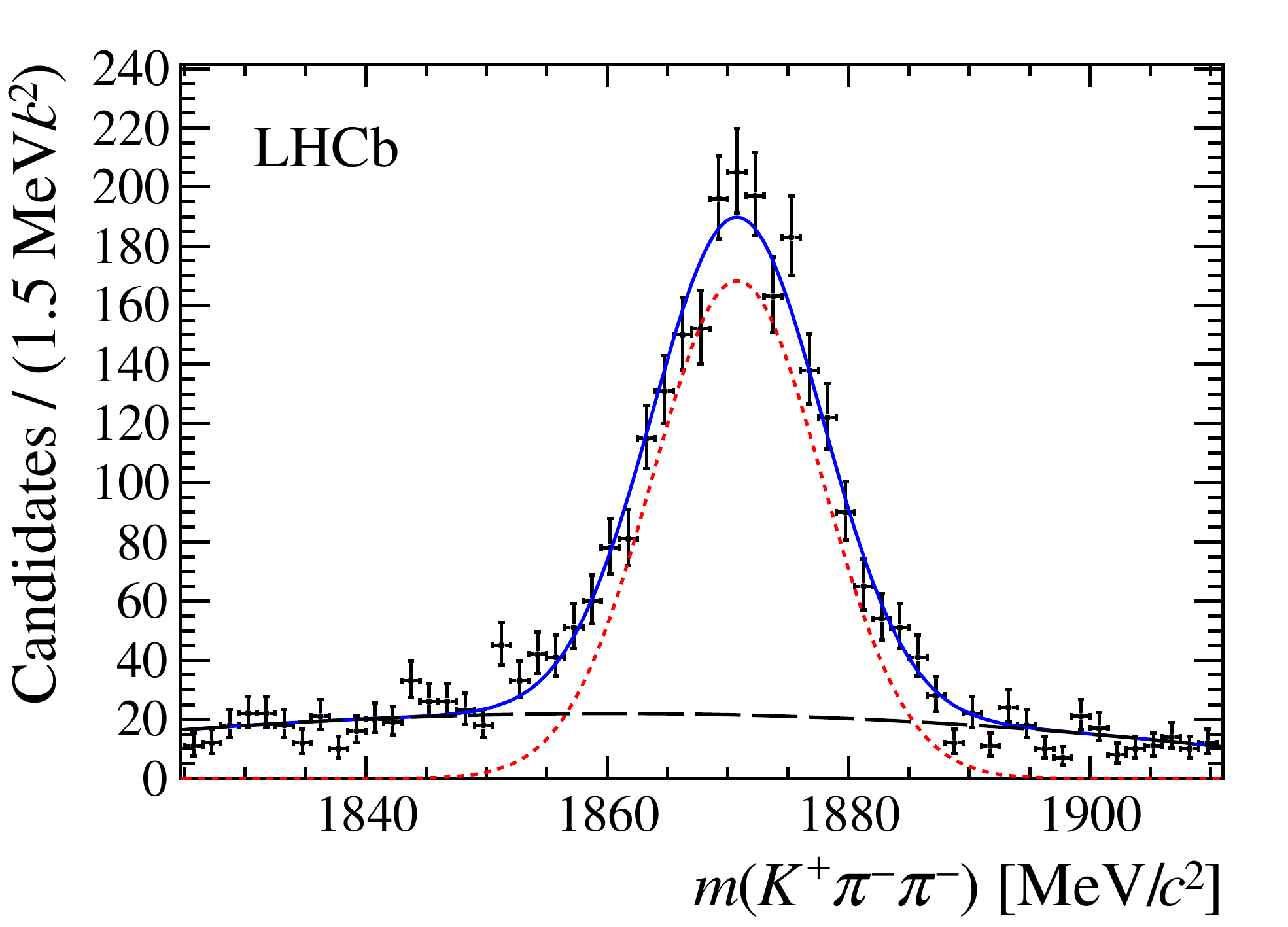}
\caption{Left: Distribution of simulated \DsEtaMuNu events as a function of dimuon mass at 8\,TeV. 
This background is removed by excluding the regions to the left of and below the black line.
Right: Fit to the 8\tev data sidebands for the bin $0.61 < \gl < 0.70$ and $0 < \pid < 0.40$ under the $K^+\pi^-\pi^-$ mass hypothesis in the \tmmm analysis. 
The short-dashed and long-dashed lines indicate Gaussian and Chebychev polynomial components. These \gl and \pid bins are not used in the analysis.}
\label{fig:bkg}
\end{figure}

\begin{figure}[t]
\centering
\begin{minipage}[b]{0.45\textwidth}
        \begin{overpic}[width=\textwidth]{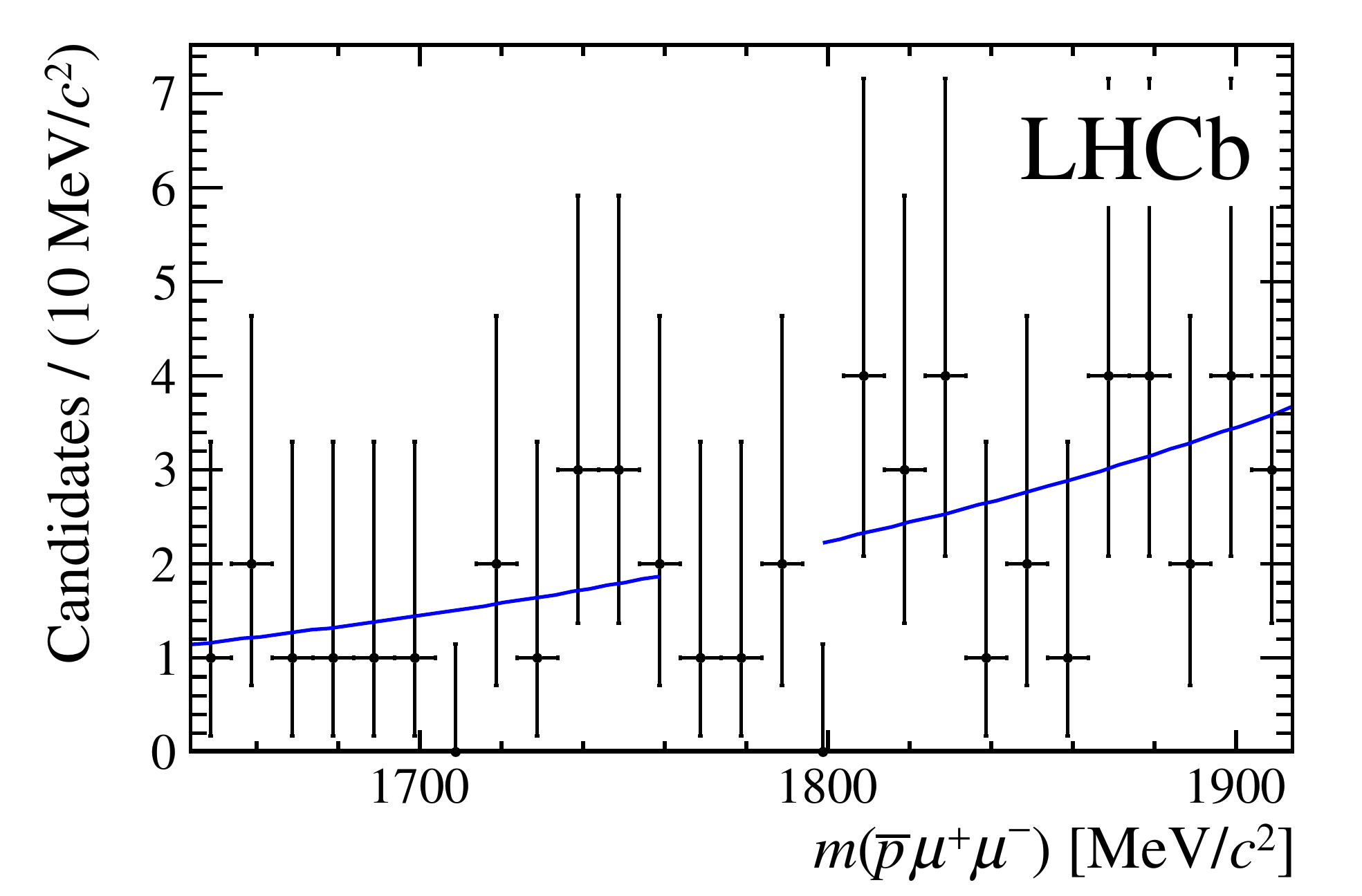}
        \put (17,55) {\footnotesize{\gl$\in [0.4, 1.0]$}}
        \end{overpic}
\end{minipage}
\begin{minipage}[b]{0.45\textwidth}
        \begin{overpic}[width=\textwidth]{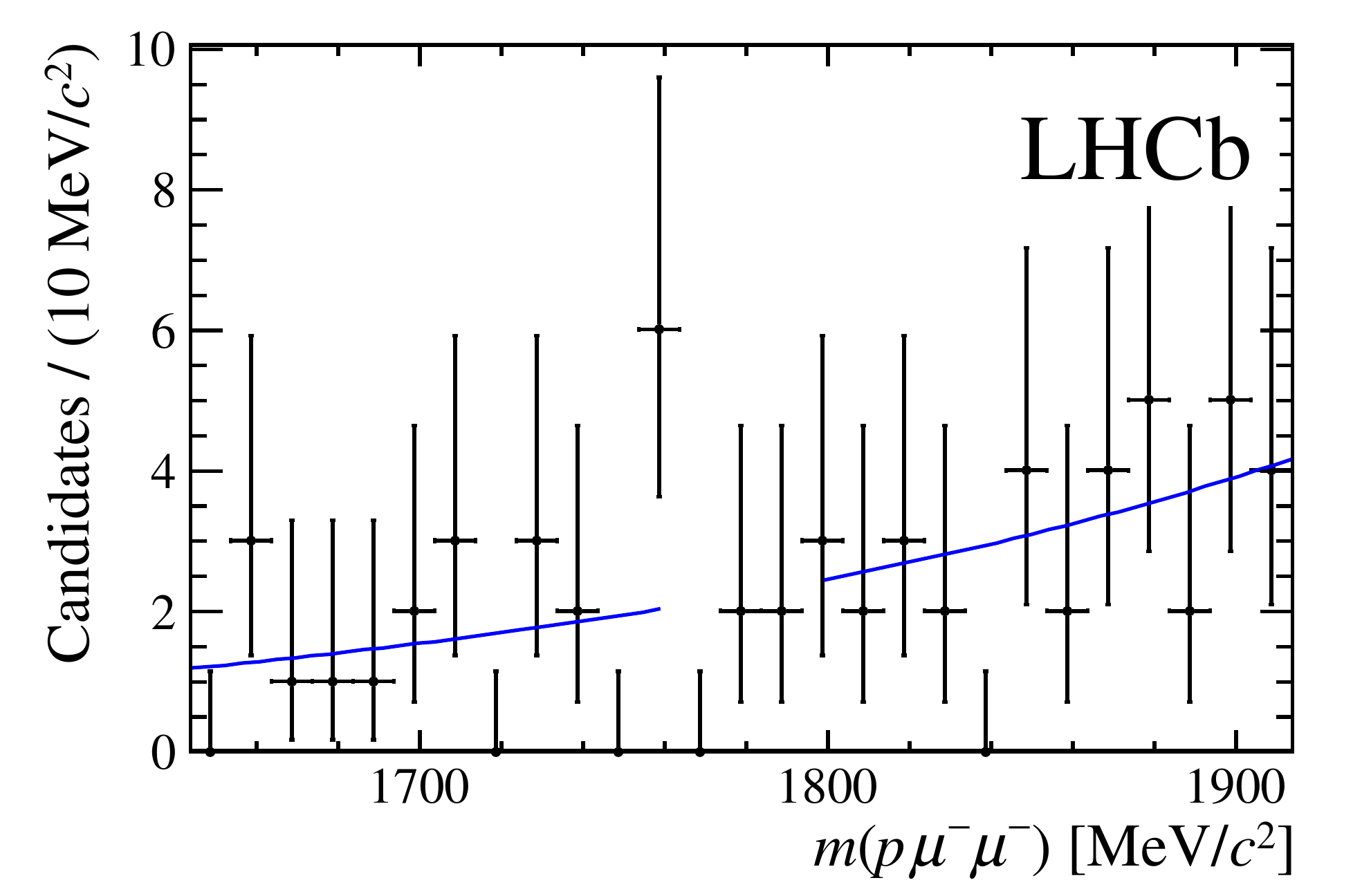}
        \put (17,55) {\footnotesize{\gl$\in [0.4, 1.0]$}}
        \end{overpic}
\end{minipage}
\begin{minipage}[b]{0.45\textwidth}
        \begin{overpic}[width=\textwidth]{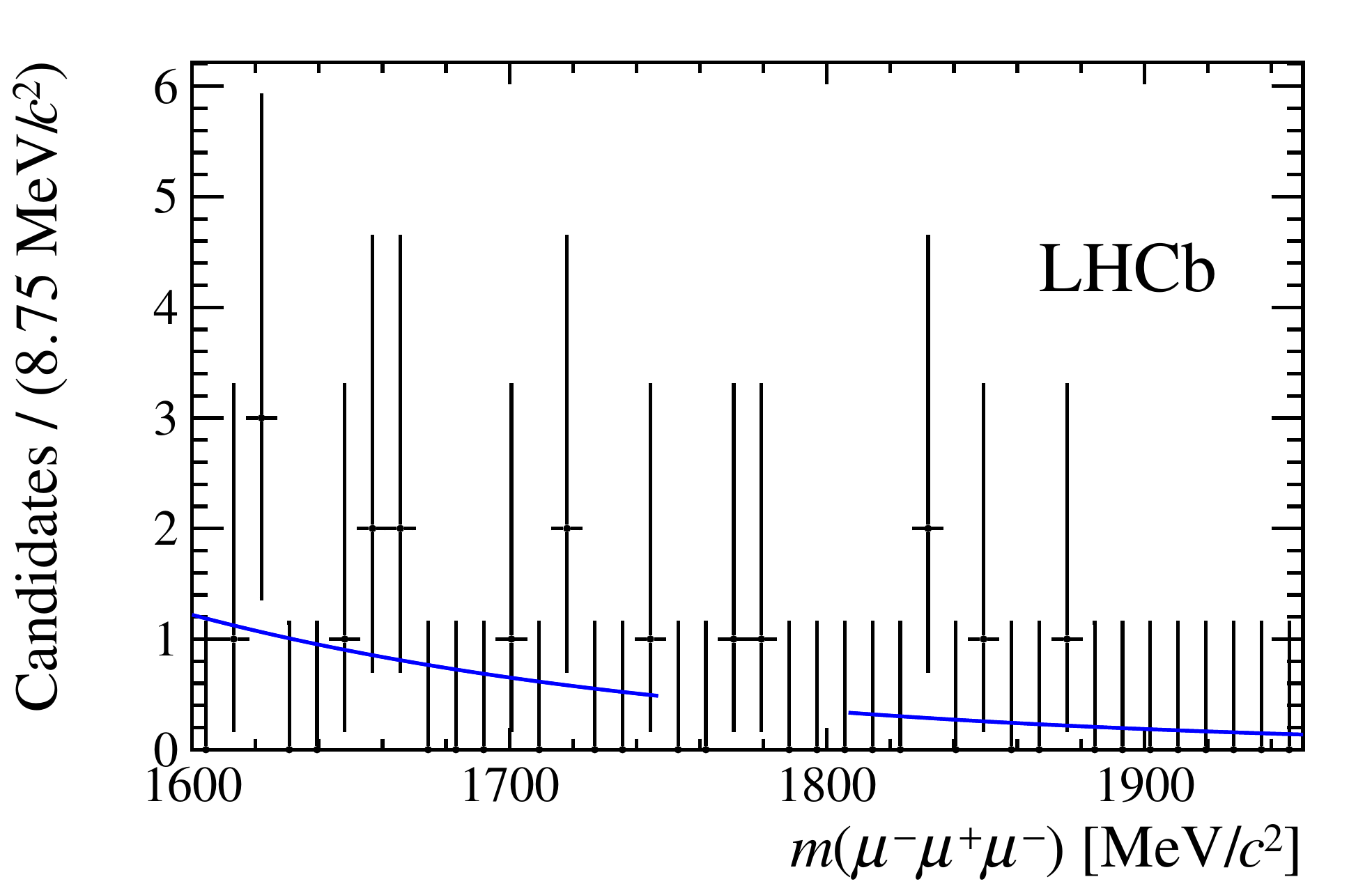}
        \put (35,55) {\footnotesize{\gl$\in [0.94, 1.0]$}}
        \put (35,50) {\footnotesize{\pid$\in [0.80, 1.0]$}}
        \end{overpic}
\end{minipage}
\caption{Fits to the mass spectra observed in the highest bins of \gl for \tpmmOS (top left), \tpmmSS (top right) and in the highest bins of \gl and \pid for \tmmm (bottom).}
\label{fig:fits}
\end{figure}

The expected number of background events per bin is calculated from an extended, unbinned maximum likelihood fit to the
mass spectrum, excluding the signal region.
An exponential function is used as the background probability density function for all three decays. 
For \tmmm the most relevant physical background comes from \DsEtaMuNu decays. About 90\% of this background is removed by requiring 
opposite-sign dimuon masses to be greater than 450\,MeV/$c^{2}$, as shown in Figure~\ref{fig:bkg} (left). 
Backgrounds from misidentified particles such as those from $D_{(s)}^-\to K^+\pi^-\pi^-$, shown in Figure~\ref{fig:bkg} (right), 
populate mainly the region of low \pid response and are reduced to a negligible level by the exclusion of the first bin of this likelihood.
From background studies no peaking backgrounds are expected in the signal windows for \tpmmOS or \tpmmSS.
Fits to the invariant mass distributions in the highest likelihood bins in the data are shown in Figure~\ref{fig:fits}.
For \tmmm in the final limit calculation the lowest bins in \gl and \pid are excluded as these are found not to contribute to the sensitivity.
 
The signal region is then unblinded and using the CL$_\mathrm{s}$ method~\cite{read, junk} we set observed limits at 90\% (95\%) CL of

\begin{eqnarray*}
\BF (\tmmm) &<& 4.6 (5.6) \times 10^{-8}, \\
\BF (\tpmmOS) &<& 3.4 (4.5) \times 10^{-7},\\
\BF (\tpmmSS) &<& 4.6 (6.0) \times 10^{-7}.
\end{eqnarray*}

These results are the first lepton flavour violation measurements at a hadron collider.  
The limits for \tmmm supercede those of Ref.~\cite{tmmmOld} and, in combination with results from the $B$ factories, improve the constraints
placed on the parameters of a broad class of NP models~\cite{hfag}. The limits for \tpmmOS and \tpmmSS represent the first ever constraints on these channels.

\section{Inclusive $Z\to\tau^+\tau^-$ cross-section at 7\,TeV}

Measurements of the $Z$ boson production cross-section are important tests of the SM. LHCb's unique coverage of the forward region allows for 
complementary results to those obtained from the general purpose detectors (GPDs) at the LHC. LHCb has previously published results using $e^+e^-$~\cite{ee} and 
$\mu^+\mu^-$~\cite{mumu} final-states. Here, measurements of the $\tau^+\tau^-$ final-state~\cite{tautau} are described. These provide a verification of previous results and a test of the universal coupling of the $Z$ to different flavours of lepton.

\begin{figure}[!t]
\centering
\includegraphics[width=0.35\textwidth]{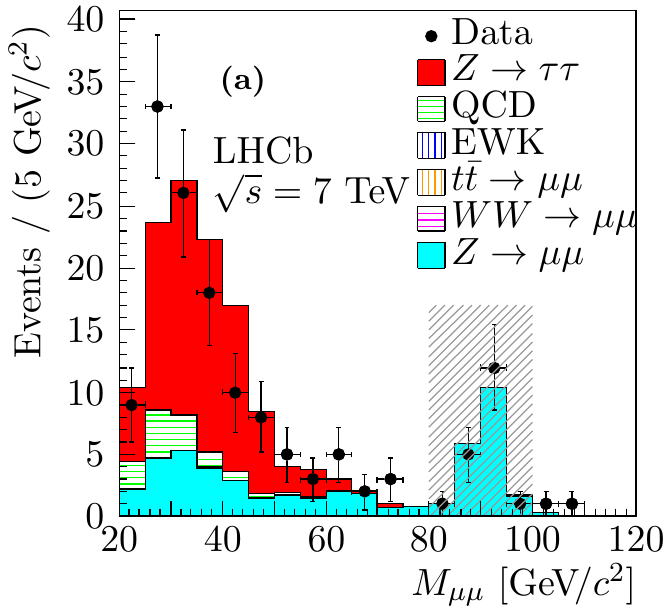}
\includegraphics[width=0.55\textwidth]{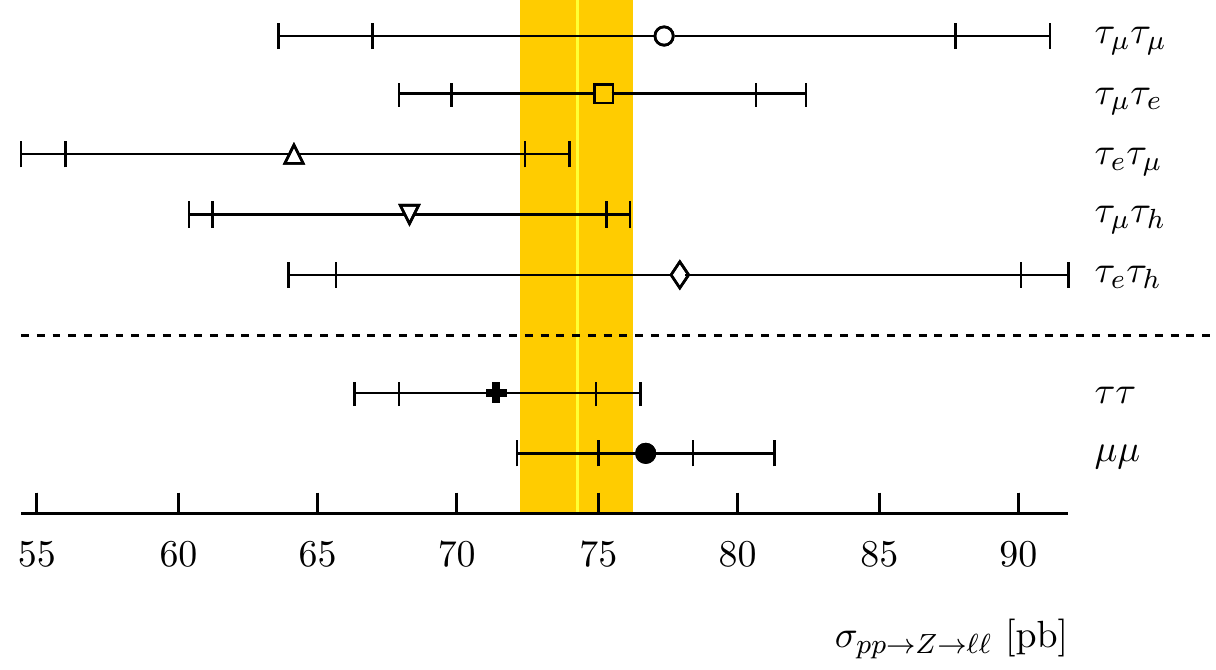}
\caption{Left: Invariant mass distribution for the $\tau_{\mu}\tau_{\mu}$ stream. The diagonal shaded region indicates the $Z\to\mu^+\mu^-$ signal peak 
which is used to create a template for the component in the $Z\to\tau^+\tau^-$ signal region ($<$ 80\,GeV/$c^2$).
Right: Cross-sections for the $Z$ decaying to the five streams defined in this analysis (open points) compared with theory (yellow band) 
and the combined $Z\to\tau^+\tau^-$ and LHCb $Z\to\mu^+\mu^-$ measurements (closed points). The inner error bars represent
the statistical uncertainties while the outer error bars represent the total uncertainties.}
\label{fig:Z}
\end{figure}

Using a sample of proton-proton collisions collected at $\sqrt{s} = 7\tev$ in 2011, which corresponds to an integrated luminosity of 1.0\,fb$^{-1}$, 
the analysis is split into five streams. These streams are denoted by $\tau_{\mu}\tau_{\mu}$, 
$\tau_{\mu}\tau_{e}$,  $\tau_{e}\tau_{\mu}$, $\tau_{\mu}\tau_{h}$ and $\tau_{e}\tau_{h}$, depending on whether the $\tau$ leptons decay to muons, electrons or hadrons, and require the first lepton to have $p_{T} >$ 20\,GeV/$c$ and 
have additional kinematic and particle identification requirements specific to each stream. Figure~\ref{fig:Z} (left) shows the invariant mass distribution 
for the $\tau_{\mu}\tau_{\mu}$ stream. The QCD, electroweak, and $Z\to\mu^+\mu^-$ backgrounds are estimated from data, whilst the $t\overline{t}$ and $W^+W^-$ backgrounds are estimated from simulation and are not visible.
	
The cross-section measurements from the individual streams are combined with the BLUE method~\cite{blue} to give 
\begin{eqnarray*}
\sigma_{pp\to Z\to\tau\tau} = 71.4 \pm 3.5 \pm 2.8 \pm 2.5\,\rm{pb},
\end{eqnarray*}
in the region of $2.0 \leq \eta^{\tau} \leq 4.5$ with $p_T^{\tau} >$ 20\,GeV/$c$ and $60 < M_{\tau\tau} <$ 120\,GeV/$c^2$. 
The first uncertainty is statistical, the second is systematic, and the third is due to the uncertainty on the integrated luminosity.

Figure~\ref{fig:Z} (right) shows a graphical summary of the individual final-state measurements, the combined measurement,
the $Z\to\mu^+\mu^-$ measurement of Ref.~\cite{mumu}, and a theory prediction. 
The theory calculation uses \texttt{DYNNLO}~\cite{dynnlo} with the MSTW08 next-to-next-leading-order parton distribution functions set~\cite{mstw}.
The results are all in good agreement.

The ratio of the combined cross-section to the LHCb $Z\to\mu^+\mu^-$ cross-section measurement is found to be
\begin{eqnarray*}
\frac{\sigma_{pp\to Z\to\tau\tau}}{\sigma_{pp\to Z\to\mu\mu}} = 0.93 \pm 0.09,
\end{eqnarray*}
and is consistent with lepton universality. The uncertainty is the combination of statistical, systematic, and luminosity uncertainties of the two measurements.



\end{document}